\documentclass[10pt,conference]{IEEEtran} 
\IEEEoverridecommandlockouts
\usepackage{cite}
\usepackage{amsmath,amssymb,amsfonts}
\usepackage{graphicx}
\usepackage{textcomp}
\usepackage{xcolor}
\def\BibTeX{{\rm B\kern-.05em{\sc i\kern-.025em b}\kern-.08em
    T\kern-.1667em\lower.7ex\hbox{E}\kern-.125emX}}

\usepackage[normalem]{ulem}
\usepackage[utf8]{inputenc}
\usepackage{color}
\usepackage [hyphens]{url}
\usepackage{algorithm}
\usepackage{algorithmicx}
\usepackage{algpseudocode}
\usepackage{braket}
\usepackage{graphicx}
\usepackage{booktabs}
\usepackage{multirow}
\usepackage{caption}
\usepackage{subfig}
\usepackage{outlines}
\usepackage{makecell}
\usepackage[flushleft]{threeparttable}

\usepackage[export]{adjustbox}
\usepackage{tikz}
\usetikzlibrary{positioning}
\usetikzlibrary{quantikz}

\long\def\comment#1{}

\newcommand*\rot{\rotatebox{90}}

\def\BibTeX{{\rm B\kern-.05em{\sc i\kern-.025em b}\kern-.08em
    T\kern-.1667em\lower.7ex\hbox{E}\kern-.125emX}}

\title{Tackling the Qubit Mapping Problem with Permutation-Aware Synthesis}

\begin{document}

\author{\IEEEauthorblockN{Ji Liu\IEEEauthorrefmark{1}$^{,\dagger}$, Ed Younis\IEEEauthorrefmark{3}$^{,\dagger}$, Mathias Weiden\IEEEauthorrefmark{4}, Paul Hovland\IEEEauthorrefmark{1}, John Kubiatowicz\IEEEauthorrefmark{4}, Costin Iancu\IEEEauthorrefmark{3}}
\IEEEauthorblockA{\IEEEauthorrefmark{1} Mathematics and Computer Science Division, Argonne National Laboratory}
\{ji.liu, hovland\}@anl.gov
\IEEEauthorblockA{\IEEEauthorrefmark{3} Computational Research Division, Lawrence Berkeley National Laboratory\\
\{edyounis, cciancu\}@lbl.gov
}
\IEEEauthorblockA{\IEEEauthorrefmark{4}Department of Electrical Engineering and Computer Science, University of California, Berkeley\\
\{mtweiden, kubitron\}@cs.berkeley.edu
}
\IEEEauthorblockA{$^\dagger$ Contributed equally to this work}
}

\maketitle
\thispagestyle{plain}
\pagestyle{plain}


\begin{abstract}
\label{sec:abs}
We propose a novel hierarchical qubit mapping and routing algorithm. First, a circuit is decomposed into blocks that span an identical number of qubits. In the second stage permutation-aware synthesis (PAS), each block is optimized and synthesized in isolation. In the third stage a permutation-aware mapping (PAM) algorithm maps the blocks to the target device based on the information from the second stage. Our approach is based on the following insights: (1) partitioning the circuit into blocks is beneficial for qubit mapping and routing; (2) with PAS, any block can implement an arbitrary $input \rightarrow output$ qubit mapping  (e.g., $q0 \rightarrow q1$) that reduces the gate count; and (3) with PAM, for two adjacent blocks we can select input-output permutations that optimize each block together with the amount of communication required at the block boundary. Whereas existing mapping algorithms preserve the original circuit structure and only introduce ``minimal''  communication via inserting SWAP or bridge gates, the PAS+PAM approach can additionally change the circuit structure and take full advantage of hardware-connectivity. 
Our experiments show that we can produce better-quality circuits than existing mapping algorithms or commercial compilers (Qiskit, TKET, BQSKit) with maximum optimization settings. For a combination of benchmarks we produce circuits shorter by up to $68\%$ ($18\%$ on average) fewer gates than Qiskit, up to $36\%$ ($9\%$ on average) fewer gates than TKET, and up to $67\%$ ($21\%$ on average) fewer gates than BQSKit. Furthermore, the approach scales, and it can be seamlessly integrated into any quantum circuit compiler or optimization infrastructure.
  
\end{abstract}
\section{Introduction}
\label{sec:intro}

Two of the most important goals of quantum compilers are circuit depth and
gate count reduction, since these are direct indicators of program
performance.  Publicly available compilers, such as Qiskit, TKET, and BQSKit, employ a sequence of gate optimization and mapping passes: (1) optimizations delete redundant gates by using  functional
equivalence~\cite{haner2018assertion_based_optimization, liu2021relaxedpeephole, kak} or pattern rewriting \cite{tket, pattern_matching}
heuristics; and (2) mapping transforms an input circuit, which may contains multiqubit gates between qubits that are not physically connected, into a circuit that can directly run on the target quantum processing unit. The qubit mapping problem is known to be NP-hard~\cite{routing_NPhard}. Several heuristic mapping algorithms~\cite{li2019sabre,   nassc, routing_bridge} as well as several optimal mappers~\cite{olsq, routing_maxsat, BIP} have been proposed. 

While ``optimizations'' delete gates, mapping introduces
additional gates to perform communication (SWAP) between qubits that are not
directly connected. 
Most existing algorithms consider only a pair of qubits as end points at any given time
and introduce 2-qubit entangling
gates (e.g., CNOT, iSWAP)
between these. A canonical representative
of such approaches is SABRE~\cite{li2019sabre}. SABRE divides the circuit into multiple layers and iteratively routes the gates in the front layer. It selects the best route based on a heuristic cost function that considers the distance between mapped physical qubits.

Topology-aware synthesis algorithms also satisfy the need for mapping a logical circuit to a physical device with limited connectivity. Based on the unitary representation of a circuit, topology-aware synthesis algorithms~\cite{qsearch, qfast, leap} generate a circuit that is compatible with a device's layout. Since synthesis algorithms directly generate a circuit based on the unitary representation, they are able to generate circuits with fewer gates than routing algorithms can. However, synthesis algorithms have scalability issues due to the exponential growth in the search space. 

In this paper we present a novel circuit mapping approach based on a hierarchical circuit representation. Our proposed framework combines  circuit synthesis with qubit mapping and routing algorithms. A circuit is first partitioned into smaller blocks. Next, we use a novel permutation-aware synthesis (PAS) to synthesize the blocks with different input and output permutations. With PAS we can optimize each block as well as find the permutation that minimizes the routing cost. Then, we use the permutation-aware mapping (PAM) framework to map and route the blocks to the target device. Integrating the synthesis algorithm in a hierarchical
mapping algorithm provides both quality and scalability for
our framework. 

\begin{figure*}
    \centering
    \subfloat[Original circuit
    \label{subfig:qft_orign}]{      \begin{adjustbox}{width=0.23\linewidth}
        \begin{quantikz}[column sep=0.1cm]
        \lstick{$q_0$}&\gate{U_3} & \ctrl{1} &\qw & \ctrl{1} & \qw & \qw & \qw & \qw & \gate{U_3} & \ctrl{2} & \qw & \ctrl{2} & \qw & \qw \rstick{$q_0$}\\
        \lstick{$q_1$}& \qw & \targ{} & \gate{U_3} & \targ{} & \gate{U_3} & \ctrl{1} & \qw & \ctrl{1} & \qw & \qw & \qw & \qw & \qw &  \qw \rstick{$q_1$}\\
        \lstick{$q_2$} & \qw & \qw & \qw & \qw & \qw & \targ{} & \gate{U_3} & \targ{} & \gate{U_3} & \targ{} & \gate{U_3} & \targ{} & \gate{U_3} & \qw\rstick{$q_2$}
        \end{quantikz}
    \end{adjustbox}
    }\hfill
    \subfloat[OLSQ\label{subfig:qft_olsq}]{ \begin{adjustbox}{width=0.23\linewidth}
        \begin{quantikz}[column sep=0.1cm]
        \lstick{$q_0$}&\gate{U_3} & \ctrl{1} &\qw & \ctrl{1} & \qw & \qw & \qw & \qw & \ctrl{1}\gategroup[2,steps=3,style={dashed,rounded corners, inner xsep=0.2pt},background]{{SWAP}} & \targ{} & \ctrl{1 } & \qw & \qw & \qw & \qw & \qw &  \qw  \rstick{$q_1$}\\
        \lstick{$q_1$}& \qw & \targ{} & \gate{U_3} & \targ{} & \gate{U_3} & \ctrl{1} & \qw & \ctrl{1} & \targ{} & \ctrl{-1} & \targ{} & \gate{U_3} & \ctrl{1} & \qw & \ctrl{1} & \qw & \qw \rstick{$q_0$}\\
        \lstick{$q_2$} & \qw & \qw & \qw & \qw & \qw & \targ{} & \gate{U_3} & \targ{} & \qw & \qw &  \qw & \gate{U_3} & \targ{} & \gate{U_3} & \targ{} & \gate{U_3} & \qw \rstick{$q_2$}
        \end{quantikz}
    \end{adjustbox}
    }\hfill
    \subfloat[Qsearch\label{subfig:qft_qsearch}]{
    \begin{adjustbox}{width=0.23\linewidth}
        \begin{quantikz}[column sep=0.1cm]
        \lstick{$q_0$}&\gate{U_3} & \ctrl{1} & \gate{U_3} & \qw & \qw & \ctrl{1} & \gate{U_3} & \ctrl{1} & \gate{U_3} & \qw & \qw & \ctrl{1} & \gate{U_3} & \qw \rstick{$q_0$}\\
        \lstick{$q_1$}& \gate{U_3} & \targ{} & \gate{U_3} & \ctrl{1} & \gate{U_3} & \targ{} & \gate{U_3} & \targ{} & \gate{U_3} & \ctrl{1} & \gate{U_3} & \targ{} & \gate{U_3} & \qw  \rstick{$q_1$}\\
        \lstick{$q_2$} & \gate{U_3} & \qw & \qw & \targ{} & \gate{U_3} & \qw & \qw & \qw & \qw & \targ{} & \gate{U_3} & \qw & \qw & \qw \rstick{$q_2$}
        \end{quantikz}
    \end{adjustbox}
    }\hfill
    \subfloat[PAS\label{subfig:qft_PAS}]{  \begin{adjustbox}{width=0.23\linewidth}
        \begin{quantikz}[column sep=0.1cm]
        \lstick{$q_0$}&\gate{U_3} & \ctrl{1} & \gate{U_3} & \qw & \qw & \ctrl{1} & \gate{U_3} & \qw & \qw & \ctrl{1} & \gate{U_3} & \qw \rstick{$q_1$}\\
        \lstick{$q_1$}& \gate{U_3} & \targ{} & \gate{U_3} & \ctrl{1} & \gate{U_3} & \targ{} & \gate{U_3} & \ctrl{1} & \gate{U_3} & \targ{} & \gate{U_3} & \qw  \rstick{$q_0$}\\
        \lstick{$q_2$} & \gate{U_3} & \qw & \qw & \targ{} & \gate{U_3} & \qw & \qw & \targ{} & \gate{U_3} & \qw & \qw & \qw \rstick{$q_2$}
        \end{quantikz}
    \end{adjustbox}
    }
    \caption{3-qubit quantum Fourier transform (QFT)  mapped to a linear topology with different algorithms}
    \label{fig:connectivity_map}
\end{figure*}
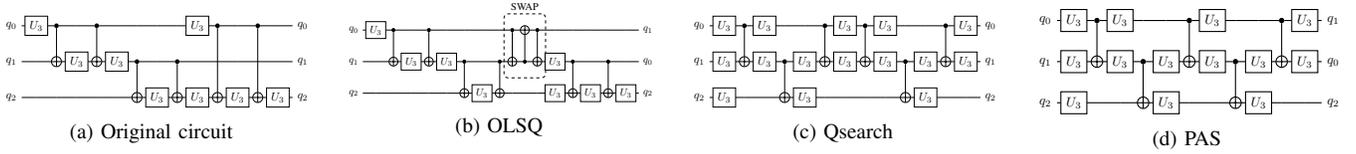

We make the following contributions:
\begin{itemize}
    \item We introduce the idea of permutation awareness and propose Permutation-Aware Synthesis (PAS). The principle behind PAS is that considering  arbitrary input-output qubit permutations at the unitary level leads to shorter circuits. These permutations are handled during post-processing without introducing any extra gates. 
    \item We present Permutation-Aware Mapping (PAM), a novel hierarchical qubit mapping framework. PAM exploits the optimization and mapping potential of PAS with block level routing heuristics. PAM has better solution quality and scalability than the optimal solver OLSQ.
    \item We demonstrate the ability to leverage hardware-connectivity. This is particularly beneficial on fully-connected architectures, such as trapped ion. In contrast, no other compiler can effectively leverage the all to all connectivity when starting from a sparsely-connected input circuit.
    
    \end{itemize}

The evaluation shows that PAM achieves  better results than state-of-the-art  available compilers: Qiskit, TKET, BQSKit. Our generated circuits contain fewer gates than optimal solutions generated by domain specific compilers for circuits such as the QFT~\cite{qft_orign} and Transverse Field Ising Model~\cite{tfim} (TFIM). PAM produces better quality results than optimal mapping algorithms such as OLSQ~\cite{olsq}, while improving scalability from less than ten qubits to thousands of qubits.

\section{Background and Motivation}
\label{sec:back}




\subsection{Qubit mapping and routing}
Qubit mapping and routing are  important  in the optimization workflow of quantum circuit compilers. 
The goal is to produce a circuit with multiqubit gates only between physically connected qubits. The problem can be resolved in two steps: finding the initial logical-to-physical qubit mapping (mapping) and applying SWAP gates to move the qubits to physically connected qubits (routing). The qubit mapping and routing problems are known to be  NP-hard~\cite{routing_NPhard}. Previous qubit mapping algorithms can be classified into two categories: heuristic algorithms and optimal mapping algorithms.

\subsubsection{Heuristic Algorithms}
SABRE~\cite{li2019sabre} is a canonical heuristic algorithm that has been adopted by the Qiskit compiler~\cite{qiskit} and multiple routing algorithms~\cite{niu2020hardware, nassc}. In SABRE, the circuit is divided into layers. The algorithm routes gates in the front layer and selects a path using a heuristic cost function based on the distance between mapped physical qubits. The heuristic cost function routes the front layer with lookahead. It balances the routing cost for the gates in the front layer and the gates in the extend layer (i.e., gates that will be routed in the future). The initial mapping is updated based on the reverse traversal of the circuit. 

Several heuristic algorithms are inspired by SABRE. Niu et al. proposed a layered hardware-aware heuristic~\cite{niu2020hardware} based on calibration datas. Liu et al.~\cite{nassc} proposed an optimization-aware heuristic that minimizes the number of 2-qubit gates after circuit optimizations. Other heuristic algorithms include TKET~\cite{tket}, commutation-based routing~\cite{routing_bridge}, simulated annealing-based routing~\cite{routing_annealing}, dynamic lookahead~\cite{routing_dynamic_lookahead}, and time-optimal mapping~\cite{routing_time_optimal}.

\subsubsection{Optimal algorithms}
Another class of routing algorithms is optimal qubit mappers. Optimal mappers solve the mapping problem by converting it into a set of constraints and finding the circuit with optimal SWAP gate count or optimal depth via optimal solvers. For example, the OLSQ~\cite{olsq} approach formulates the problem as a satisfiability modulo theories (SMT) optimization problem and then uses the Z3 SMT solver~\cite{z3solver} to find the optimal circuit. The BIP mapper~\cite{BIP} in Qiskit finds the optimal mapping and routing by solving a binary integer programming (BIP) problem. Because of the exponential growth of the search space, however, these constraint-based solvers all face significant scalability issues. Besides the scalability issue, we will show in the next subsection that synthesis algorithms may generate smaller circuits than those that the optimal mappers generate.

Mapping and routing should be performed at the multiqubit gate level. The Orchestrated Trios~\cite{Trios} compiler shows that preserving complex 3-qubit operations during qubit mapping and routing can reduce routing overhead, since these complex operations better inform the routing algorithm about how qubits should be moved around. However, a generalized quantum algorithm may not contain these complex operations. In the permutation-aware mapping (PAM) framework, we propose a block-based routing framework that captures circuit structure and has a better lookahead window. \textbf{Insight 1: Creating and preserving complex operations (blocks) may be beneficial for qubit mapping and routing.}

Both heuristic and optimal solvers  introduce communications by inserting SWAP gates or bridge gates, but they are unable to remove communication. Another advantage of  block-based mapping is the ability to take permutation into consideration. A block with different input and output permutations may have different basis gate counts. We can find the input and output permutations that optimize each block. These permutations at the block boundary can be resolved as we map the circuit to physical devices. \textbf{Insight 2: Introducing input and output permutation for the circuit block may reduce the circuit size.}

\begin{figure*}
    \centering
    \includegraphics[width=0.8\textwidth, height = 8.5cm]{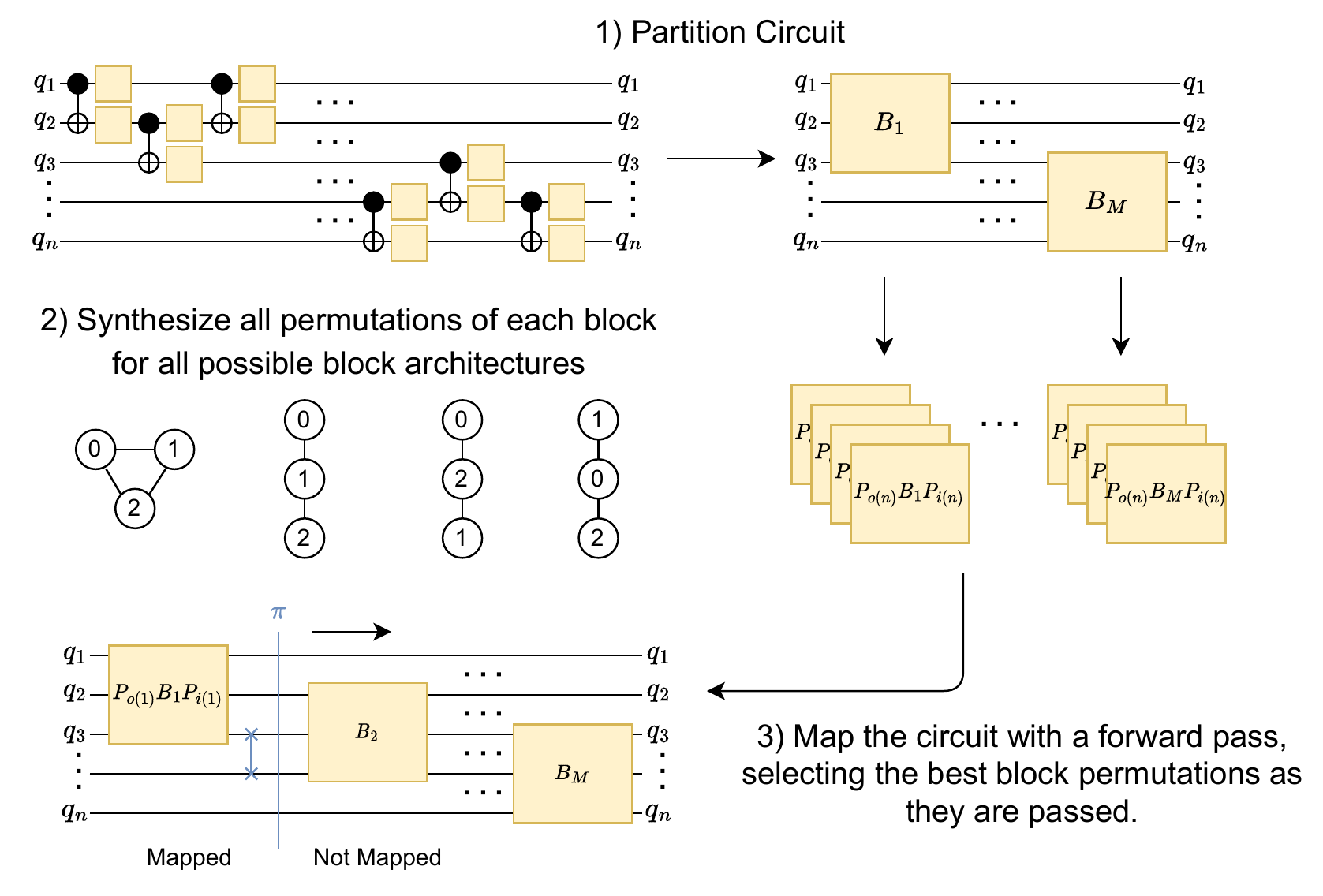}
    \caption{\footnotesize \it
    This example applies permutation-aware mapping on a circuit with
    3-qubit blocks. Each block is synthesized with possible input and output permutations. Our permutation-aware mapping
    procedure then resolves the different qubit permutations as we map the blocks to the device. Additionally, as with any
    mapping or routing algorithm, inserting SWAP gates is necessary.
    }
    \label{fig:pam}
\end{figure*}

\subsection{Synthesis for mapping}
\label{subsec:syn_for_mapping}

A unitary synthesis algorithm generates a quantum circuit based on a $2^n \times 2^n$ unitary matrix representation of the circuit. In this paper we focus on topology-aware synthesis algorithms since they can synthesize quantum circuits that are compatible with the device topology. Qsearch~\cite{qsearch} is a topology-aware synthesis algorithm that generates circuits with optimal depth. Qsearch employs an A* heuristic to search over a tree of possible circuits based on device topology. In practice, however, the scalability of Qsearch  is limited to 4 qubits because of the exponential growth of the search space. Other topology-aware synthesis algorithms such as QFAST~\cite{qfast} improve the scalability of synthesis by encoding placement and topology using generic “gates.” 
LEAP~\cite{leap} improves scalability from 4- to 6-qubit circuits by narrowing the search space through prefix circuit synthesis. 
The QGo~\cite{qgo} optimization framework proposes a circuit partitioning algorithm that partitions a large circuit into smaller blocks (subcircuits) and synthesizes each block in parallel. In contrast to our work, QGo is an optimization framework and is applied after qubit mapping and routing. It relies on the qubit mapping algorithm to find the logical-to-physical qubit mapping and cannot change the placement of the blocks. 

Synthesis algorithms can map small circuits to physical devices with fewer gates compared with  routing algorithms. We use the 3-qubit QFT algorithm as an example. The best-known implementation of this algorithm~\cite{qft_orign} is shown in Figure~\ref{subfig:qft_orign} which contains six CNOT gates. As shown in Figure~\ref{subfig:qft_olsq}, the circuit mapped with optimal routing algorithm OLSQ~\cite{olsq} contains a single SWAP gate, and the total CNOT gate count is nine. Since  routing algorithms only insert SWAP gates, they maintain the original structure of the circuit. The subsequent optimizations may not change these structures and hence may result in a suboptimal solution. 

On the other hand,  synthesis algorithms directly construct a circuit based on the unitary matrix, regardless of the original circuit structure. In Figure~\ref{subfig:qft_qsearch}, the Qsearch algorithm finds a better linearly connected design with only six CNOT gates. The permutation-aware synthesis (PAS) further reduces the gate cost by finding the best output permutation and implements the 3-qubit QFT with only five CNOTs, shown in Figure~\ref{subfig:qft_PAS}. To the best of our knowledge, this is the best-known implementation of this essential circuit. \textbf{Insight 3: Permutation-aware synthesis can directly map a circuit to a physical device with higher quality than routing algorithms can achieve.}

    
    
\section{PAM Overview}
\label{sec:overview}

We propose our permutation-aware mapping (PAM) framework based on the aforementioned insights. PAM partitions a circuit into blocks, performs permutation-aware synthesis for each block, and runs a block-based mapping for the blocks.

An overview of the PAM framework is shown in Figure~\ref{fig:pam}. First, the input n-qubit quantum circuit is vertically partitioned into k-qubit blocks, $B_{1}$, ... , $B_{M}$, by grouping together adjacent gates. Second, we resynthesize each block for all possible permutations and k-qubit topologies (sub-topologies). As shown in the previously mentioned example, 3-qubit blocks generally need four sub-topologies. One comes from the fully-connected or all-to-all 3-qubit architecture, and the other three represent all orientations of a 3-qubit line or nearest-neighbor architecture. The resynthesis results, including the associated qubit permutation and circuit, are stored for use during the next mapping phase.

\begin{figure*}
    \centering
    \includegraphics[width=0.9\textwidth]{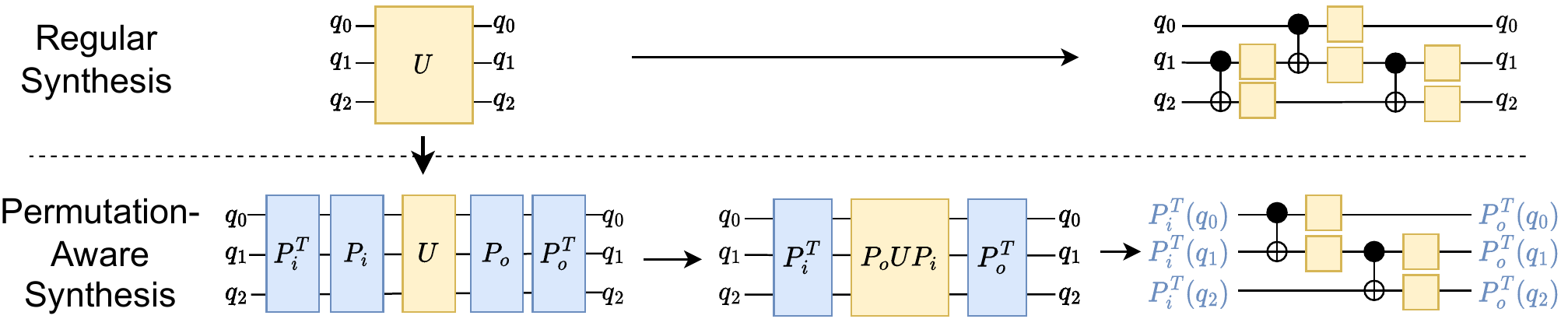}
    \caption{\footnotesize \it
        Regular synthesis will construct a circuit implementing a given unitary matrix preserving input and output qubit orderings.
        Before synthesizing a unitary, permutation-aware synthesis will
        factor out implicit qubit communication, leading to an overall shorter
        circuit. This action, however, will not preserve input and output qubit
        orderings and will require some simple classical processing when
        preparing the initial qubit state and reading out the final qubit
        state.
    }
    \label{fig:pas}
\end{figure*}

The permutation-aware mapping algorithm continues over the partitioned circuit, unlike standard heuristical mappers, which mainly deal with native gates. To accomplish this, we supplement the SABRE algorithm with a novel, albeit generalized, heuristic to evaluate the current mapping state. Additionally, we add an extra processing step when moving gates from the unmapped to the mapped region. During this step, we utilize another novel heuristic to select the synthesized block permutation that best balances gate count and routing overhead for subsequent blocks. These block-level permutations leverage implicit communication buried in their computation to beneficially affect the state of the progressing mapping algorithm, drastically reducing the need for SWAP gates to perform global communication.

\section{Permutation-Aware Synthesis}
\label{sec:pas}




\subsection{Permutation-aware synthesis}
\label{subsec:PAS_define}
Synthesis algorithms typically construct a circuit based on its unitary matrix representation. They preserve the input and the output qubit ordering for the circuit. However, the  ordering can be permuted to change the unitary matrix. Constructing the circuit based on the permuted unitary may result in a shorter circuit.

We formalize here the concept of permutation-aware synthesis. As shown in Figure~\ref{fig:pas}, a synthesis algorithm constructs the unitary matrix $U$ with three CNOTs, and the qubit ordering is preserved. When considering alternate qubit permutations, we can always insert an input order permutation $P_i$ and its inverse $P_i^{T}$. $P_i$ and $P_i^{T}$ will cancel out, and the circuit's functionality is unchanged. Similarly, we can insert an output order permutation $P_o$ and its inverse $P_o^{T}$. After introducing these four extra permutations, we can group $P_i$, $U$, and $P_o$ to generate a permuted unitary for synthesis. Since applying a gate on the left of a gate is equivalent to multiplying its unitary matrix on the right, the permuted unitary gate is represented as $P_oUP_i$. This permuted unitary gate may require fewer basic gates to implement. The insertion of permutations ${P_i}^T$ and ${P_o}^T$ can be achieved through classically changing the qubit index orders and does not require any gates to implement. As shown in Figure~\ref{fig:pas}, the gate marked in yellow is a permuted unitary gate, which can be synthesized with only two CNOT gates. The permutations ${P_i}^T$ and ${P_o}^T$ have the effect of permuting the input and output qubit orders, but can be handled classically by changing the index of the input and output qubit orders. In other words, permutations can be factored out to the inputs and outputs and resolved through classical processing. The core idea of our permutation-aware synthesis and permutation-aware mapping algorithm is the association of the original unitary with input and output permutations and the methodology to resolve the remaining permutations. We will evaluate permutations to find the one that yields the fewest gates.

\subsection{Permutation search space}
\label{subsec:pas_search_space}
Permutation-aware synthesis is a generalized approach that can be applied to any synthesis algorithm. In this work we use the Qsearch~\cite{qsearch} algorithm to synthesize the permuted unitary since it produces near-optimal solutions.

Multiple permutations need to be evaluated to identify the best permutation. For an $n$-qubit circuit, there are $n!\times n!$ input and output permutation combinations in total. To reduce the search space, we introduce the sequentially permutation-aware synthesis (SeqPAS). In SeqPAS, we first evaluate all the input permutations to find the best permutation $P_i$. Then, we fix the best input permutation $P_i$ to find the best output permutation $P_o$. The total number of evaluations in SeqPAS is $2\times n!$. In fully permutation-aware synthesis (FullPAS), we evaluate all the input and output permutation combinations. In  Section~\ref{subsec:block_mapping} we will provide a comprehensive comparison of these two PAS designs with the synthesis and routing algorithms. In most cases, SeqPAS generates circuits with a gate count close to that of FullPAS and with less compilation overhead. It's worth noting that classical reversible logic synthesis~\cite{wille2009classical_permutation_garbage, saeedi2010classical_permutation_garbage} have leveraged output permutations to reduce circuit cost, they reduce the search space with garbage output bits (the outputs are by definition don’t cares in the reversible circuit). However, in the practical quantum circuits, non of the output qubits is garbage qubit.

\section{Permutation-Aware Mapping}
\label{sec:pam}
In this section we clarify how we combine circuit partitioning, block-level permutation-aware resynthesis, and novel heuristics with tried-and-trued routing techniques to assemble our permutation-aware mapping framework. 

Like other mapping algorithms, we break the problem into two steps: layout and routing. Layout discovers an initial logical to physical qubit mapping; routing then progresses this mapping through the circuit, updating it and adding SWAP gates as necessary to connect interacting logical qubits. While these two steps are distinct in our framework, the same circuit sweep methods that utilize circuit partitions and permutation-aware synthesis implement both. As such, we first describe our partitioning and resynthesis steps and then detail our heuristic circuit sweep. After proposing the full algorithm, we provide an analysis of PAM's computational complexity.

\subsection{Circuit partitioning}
\label{subsec:Partition}

The PAM algorithm first partitions a logical circuit vertically into k-qubit blocks. Vertical partitioning groups together gates acting on nearby qubits into blocks and is commonly implemented by placing gates into bins as a circuit is swept left to right.
This method contrasts horizontal partitioning techniques~\cite{baker2020time_slice} used in distributed quantum computing to best separate qubits. The binning approach to partitioning is excellent for our algorithm due to its scalability. These partitioners are linear with respect to gate count, $O(M)$, and we found that alternative partitioning techniques showed little variance in experimental results.

\subsection{Permutation-aware resynthesis}
\label{subsec:PASblocks}

After partitioning a circuit, we represent it as a sequence of k-qubit logical blocks containing the original gates. Later, layout and routing will replace these blocks with one of many permutated versions. Having all block permutations accessible enables our heuristic to compare the quality of each and select the one that best balances its gate count with its effect on mapping. To discover all possible block permutations, we use permutation-aware synthesis. 

If we perform permutation-aware synthesis online during mapping, we would serialize the required block synthesis. For very large circuits, this will become intractable very quickly. To overcome this, we perform the resynthesis step across all blocks in parallel offline. While this is an embarrassingly parallel problem, performing it offline has the added challenge of not knowing the block's final physical position and, therefore, its required topology. As a result, we will also need to synthesize for different topologies in addition to permutations.

We now resynthesize each block once for each possible permutation and connectivity requirement. We synthesize for all topologies because it allows us to identify extra connections provided by the hardware. For example, as shown in Figure~\ref{fig:pam}, every three-qubit block will have six possible input permutations, six possible output permutations, and four different possible connectivities. Naively, this totals 144 synthesis calls for every three-qubit block.

We can dramatically reduce the number of required synthesis calls in two ways. First, we can perform a quick sub-topology check of the target architecture to eliminate possible connectivities. For example, if the target architecture is only linearly-connected, we do not need to consider the all-to-all connectivity requirement during block resynthesis. This is because no possible placement of a 3-qubit block on a linearly-connected topology can ever be fully-connected. Although not intuitive, it is common to eliminate some required sub-topologies when targetting realistic architectures with 3-qubit blocks.

The second way to reduce the number of synthesis calls is to recognize equivalent permutations. One can permute a resynthesized circuit to produce a new circuit implementing the same unitary with different input and output permutations and a rotated topology. Since there are $n!$ ways to permute a circuit, we can reduce the number of synthesis calls required for permutation-aware resynthesis by that many. After applying this optimization to 3-qubit blocks, we only need to synthesize a max of 24 different unitaries.

\subsection{Heuristic circuit sweep}
\label{subsec:RoutingHeuristic}
PAM's layout and routing algorithms utilize the same circuit sweep responsible for evolving a given logical-to-physical qubit mapping through a circuit. This section describes how we augment the SABRE algorithm~\cite{li2019sabre} to leverage block-level permutations.

We follow the SABRE convention in dividing the logical circuit into a front layer $F$ and $E$, an extended set. The front layer consists of gates with no predecessors, and the extended set consists of the first $|E|$ successors of the front layer, where $|E|$ is configurable. The extended layer $E$ is defined for lookahead analysis. As the sweep builds the physical circuit, it removes gates from the logical circuit and updates $F$ and $E$.

In our first change from the SABRE algorithm, we generalized the heuristic cost function from \cite{routing_layer} to support arbitrary-sized gates given by:

$$\mathcal{F}(\pi) = \frac{1}{|F|}\sum_{b\in F}\sum_{i,j\in b} D[\pi(b.i)][\pi(b.j)]$$
$$\mathcal{E}(\pi) = \frac{W_E}{|E|}\sum_{b\in E}\sum_{i,j\in b} D[\pi(b.i)][\pi(b.j)]$$
$$H(\pi) = \mathcal{F}(\pi) + \mathcal{E}(\pi)$$

Here $b$ is a gate block. $D$ is the distance matrix that records the distance between physical qubits. $|F|$ and $|E|$ are the size of the front and extended layers, respectively. Minimizing this heuristic requires bringing all front layer gates' logical qubits physically closer together. To add lookahead capabilities, the operations in the extended set also contribute a term weighted by a configurable value $W_E$.

It is essential to note some challenges with heuristic mapping algorithms when generalizing from two-qubit to many-qubit gates. There are many ways to bring more than two qubits together on a physical architecture, creating many local minimums in a heuristic swap search. To combat this, we disabled swaps between any pair of logical qubits if an operation exists in the front layer containing both.

The second change we make to the SABRE algorithm is adding a step when removing an executable gate from the front layer and placing it in the physical circuit. In our case, the gates are blocks, and we have already pre-synthesized their permutations. The current mapping determines the block's input permutation and sub-topology, leaving the block's output permutation to be freely chosen. For 3-qubit blocks, we will have six possible choices for output permutation.

Two factors determine which output permutation to select for a given block. The chosen permutation will alter the ongoing mapping process potentially for the better. Also, the circuits associated with each permutation will have differing gate counts. We want to choose an output permutation that balances the resulting block's gate count with the overall effect on mapping. We modify the swap search heuristic to select the best permutation, producing the following heuristic:

$$ P(\pi) = W_P\times C[b][G_b][(P_i,P_o)] + H(P_o(P_i(\pi))) $$

Here $C[b][G_b][(P_i,P_o)]$ is the 2-qubit gate count for the block $b$ with subtopology $G_b$ and permutations $(P_i, P_o)$. The $W_P$ weights the gate cost with the mapping cost and has been empirically discovered to be 0.1. Note that after applying the permuted block,
the mapping cost function is evaluated using mapping updated by both input and output permutations.

In summary, our circuit sweep iterates over a partitioned circuit inserting swaps according to a swap search with a generalized heuristic to make blocks in the front layer executable. At this point, they are moved to the physical circuit and assigned a permutation according to a novel heuristic that updates the mapping state as the algorithm advances.

\subsection{Layout and routing}
\label{subsec:PAMLayout}

Both of PAM's layout and routing algorithms are built trivially using the circuit sweep method previously described. Similar to SABRE, layout is conducted by randomly starting with an initial mapping and evolving it via the heuristic circuit sweep. Once complete, layout evolves the resulting mapping through the reverse of the logical circuit. This back-and-forth process is repeated several times until a stable mapping has been discovered. Routing then performs a single forward pass of the circuit sweep starting from the mapping that layout found.

Some corner cases exist where the heuristic may not select the best permutation. After routing the circuit, we can catch these corner cases by repartitioning and resynthesizing the circuit. The repartitioning process will group newly placed SWAP gates with other operations. This process is termed as gate absorption in some prior works~\cite{BIP,olsq_ga}. However, these works primarily discussed the absorption of SWAP gates with SU(4) gates. In our case, repartitioning and resynthesis of many-qubit blocks and swap networks allow us to reduce circuit gate count further.

\subsection{Complexity analysis}
\label{subsec:Complexity}
The PAM framework is scalable in terms of both the number of qubits $N$ and the total 2-qubit gate count $M$. It has the same level of time complexity as SABRE, which is $O(N^{2.5}M)$.

The PAM framework consists of four  compilation steps. First, a circuit is partitioned into gate blocks with the partitioning algorithm. The default quick partition algorithm~\cite{bqskit} in BQSKit has complexity of $O(M)$. Second, we use PAS to synthesize the permutations for each block. Since we limit the block size to less than three, the synthesis time for each block is bounded by a constant time limit $O(C)$. In the worst case, the total number of block equals  the total number of gates $M$ over the constant block size. Therefore, the PAS step has time complexity of $O(M)$. The layout step and the routing step in the worst case have the same time complexity as does the SABRE routing algorithm, $O(N^{2.5}M)$. By adding all the steps together, the PAM framework has time complexity of $O(N^{2.5}M)$, which is as scalable as that of other heuristic routing algorithms.
\section{Experimental Setup}
\label{sec:exp}
The permutation-aware synthesis and mapping algorithms are implemented
by using the BQSKit framework~\cite{bqskit}. 
We compared the proposed algorithms with the original SABRE algorithm
and three industrial compilers: Qiskit~\cite{qiskit}, TKET~\cite{tket},
and BQSKit. When possible, we additionally compared the algorithms with an optimal mapping algorithm OLSQ~\cite{olsq} followed by Qiskit optimizations. 

\subsection{Benchmarks}

We used two sets of benchmarks to evaluate the proposed
permutation-aware algorithms. When evaluating algorithms at the block
level, we used a collection of small 3-, and 4-qubit circuits,
which are either commonly used as building blocks in larger quantum
programs or represent a smaller version of standard programs. These are
listed in Figure~\ref{fig:pas_benchs}. Qiskit generated all of them except
for the QAOA circuit, which was generated by Supermarq~\cite{supermarq}.
The Toffoli and Fredkin gates are well studied, and often compilers will
be able to handle them through optimized workflows. To ensure a diverse
benchmark set, we included some less-optimized gates: the singly and
doubly controlled-MS XX gate~\cite{msgate}. QFT and QAOA circuits were
included because they have been used extensively in past benchmark sets.
Supermarq~\cite{supermarq} generated the 4-qubit, fermionic-SWAP QAOA circuit.

\begin{figure}[htbp]
    \centering
    \subfloat[small block benchmarks\label{fig:pas_benchs}]{
    \adjustbox{width = 0.45\linewidth, valign=t}{
        \small
    \begin{tabular}{c|c}
        Benchmark & CNOT Gates \\
        \hline
        ccx3   & 6   \\
        cswap3 & 8   \\
        cxx3   & 22  \\
        ccxx4  & 118 \\
        qft3   & 6   \\
        qft4   & 12  \\
        qaoa4  & 18  \\
    \end{tabular}
    }
    }
    \hfill
    \subfloat[large quantum benchmarks\label{fig:pam_benchs}]{
        \adjustbox{width = 0.45\linewidth, valign=t}{
            \small
        \begin{tabular}{c|c}
            Benchmark & CNOT Gates \\
            \hline
            adder63 & 1405  \\
            mul60   & 11405 \\
            qft5    & 20    \\
            qft64   & 1880  \\
            grover5 & 48    \\
            hub18   & 3541  \\
            shor26  & 21072 \\
            qaoa12  & 198   \\
            tfim64  & 4032  \\
            tfxy64  & 4032  \\
        \end{tabular}
        }
    }
    \caption{
    \footnotesize \it
        Two-qubit gate counts for the small block and large quantum program benchmark suites.
        The number of qubits in the circuit is given as a suffix.
    }
    \label{fig:all_benchs}
\end{figure}

\begin{figure*}
    \centering
    \begin{adjustbox}{width=0.9\linewidth}
    \subfloat[CNOT counts]{
        \small
        \begin{tabular}{rrrrrrrr|}
            &   \rot{ccx} &   \rot{cswap} &   \rot{cxx} &   \rot{qft3} &   \rot{qft4} &   \rot{qaoa4}  &   \rot{ccxx}\\
            \hline
             Qiskit  &     9 &      10 &    31&      7 &     17 &      18 &    270  \\
             TKET    &     9 &      10 &    29 &      9 &     21 &      17 &    172 \\
             OLSQ+Opt    &     9 &      10 &    21 &      9 &     17 &      18  &    184\\
             Qsearch &     8 &       8 &     5 &      6 &     16 &      18 &     15 \\
             SeqPAS    &     7 &       8 &     4 &      6 &     14 &      14 &     13 \\
             FullPAS     &     7 &       8 &     4 &      5 &     13 &      12 &     10 \\
        \end{tabular}
    }
    ~
    \subfloat[Compile time in seconds]{
        \small
        \begin{tabular}{|rrrrrrrl}
            \rot{ccx} &   \rot{cswap} &   \rot{cxx} &   \rot{qft3} &   \rot{qft4} &   \rot{qaoa4} &   \rot{ccxx} \\
            \hline
             2.43 &    2.43 &  2.61 &   2.41 &     2.51 &     2.46 &     3.76 & Qiskit   \\
             0.05 &    0.07 &  0.16 &   0.07 &     0.13 &     0.15 &     0.92 & TKET     \\
             2.66 &    2.66 &  5.51 &   2.68 &     3.40 &     2.72 & 19325.55 & OLSQ+Opt     \\
            10.21 &    7.51 &  1.78 &   3.23 &    90.45 &   216.25 &    54.33 & Qsearch  \\
            23.73 &   29.45 &  7.47 &   7.98 &  5188.93 &  2705.54 &  5974.15 & SeqPAS     \\
            23.83 &   83.38 &  7.49 &   9.80 & 46733.13 & 16582.57 & 36676.70 & FullPAS      \\
        \end{tabular}
    }
    \end{adjustbox}

    \caption{\footnotesize \it
        Common quantum circuit building blocks compiled to a linear
        topology using varying methods.
    }
    \label{tab:linepas}
\vspace{-5mm}
\end{figure*}

\begin{figure*}
    \centering
    \begin{adjustbox}{width=0.9\linewidth}
    \subfloat[CNOT counts]{
        \small
        \begin{tabular}{rrrrrrrr|}
            &   \rot{ccx} &   \rot{cswap} &   \rot{cxx} &   \rot{qft3} &   \rot{qft4} &   \rot{qaoa4}  &   \rot{ccxx}\\
            \hline
             Qiskit  &     6 &       7 &    17&      6 &     12 &      18  &    114 \\
             TKET    &     6 &       7 &    17&      6 &     12 &      12  &     95 \\
             OLSQ+Opt    &     6 &       7 &    17 &      6 &     12 &      18 &    114 \\
             Qsearch &     6 &       7 &     5&      6 &     12 &      13  &     11 \\
             SeqPAS    &     6 &       7 &     5  &      6 &     13 &      12 &      9\\
             FullPAS     &     6 &       7 &     4  &      5 &     10 &       9&      9 \\
        \end{tabular}
    }
    ~
    \subfloat[Compile time in seconds]{
        \small
        \begin{tabular}{|rrrrrrrl}
            \rot{ccx} &   \rot{cswap} &   \rot{cxx} &   \rot{qft3} &   \rot{qft4} &   \rot{qaoa4} &   \rot{ccxx}\\
            \hline
             2.42 &    2.39 &  2.47 &   2.40 &      2.45 &      2.44 &     3.20 & Qiskit   \\
             0.06 &    0.07 &  0.15 &   0.07 &      0.13 &      0.15 &     0.92 & TKET     \\
             2.61 &    2.61 &  2.81 &   2.59 &      2.75 &      2.73 &     5.27 & OLSQ+Opt     \\
             7.40 &   11.34 &  2.61 &   6.45 &   1683.54 &   2752.78 &   336.55 & Qsearch  \\
            19.73 &   46.65 &  5.88 &   9.79 &   4117.27 &   1793.36 &  1232.44 & SeqPAS     \\
            25.64 &   65.95 &  8.85 &  13.46 & 234174.05 & 101642.08 & 20775.23 & FullPAS      \\
        \end{tabular}
    }
    \end{adjustbox}

    \caption{\footnotesize \it
        Common quantum circuit building blocks compiled to a fully-connected
        topology using varying methods.
    }
    \label{tab:atapas}
    \vspace{-5mm}
\end{figure*}

To evaluate the qubit mapping and circuit optimization capabilities of
our proposed algorithm against full-scale compilers, we used a benchmark
suite consisting of 10 real quantum programs of various types ranging
in size from 5 to 64 qubits. We included two commonly used arithmetic
circuits~\cite{ripple_adder, qgo}, which contain long chains
of 2-qubit gates. These chains are worst-case scenarios for partitioning
compilers and are useful to evaluate. We included a 5-qubit Grover and
26-qubit Shor circuit generated by Qiskit~\cite{grover, shor1999}.
The suite also included two variational quantum algorithms: 
Supermarq's 12-qubit fermionic-SWAP QAOA circuit~\cite{supermarq, qaoa} and
 an 18-qubit circuit simulating a spinful Hubbard model
generated with OpenFermion~\cite{hubbard, openfermion}. Moreover, we included
two real-time evolution circuits: a transverse-field ising
(TFIM)~\cite{tfim}
and a transverse-field XY (TFXY) model.
The constant-depth F3C++ compiler~\cite{constant_depth, camps2022algebraic_f3c++, kokcu2022algebraic_f3c++} produced these
circuits, which before PAM were the best implementations.
Figure~\ref{fig:pam_benchs} lists all large quantum program benchmarks alongside their gate
counts.

\subsection{Experiment platform}
All experiments were executed with Python 3.10.7 on a 64-core AMD Epyc
7702p system with 1 TB of main memory running Ubuntu 20.04 as the
operating system. We used versions 1.0.3, 0.38.0, 1.6.1, and 0.0.4.1 for
the BQSKIT, Qiskit, PyTKET, and OLSQ packages, respectively.

\subsection{Algorithm configuration}
Unless otherwise specified, we used the Qsearch algorithm for 3-qubit
synthesis and the LEAP algorithm for 4-qubit synthesis. For both, we
used the BQSKit implementation configured with the recommended settings:
4 multistarts and the default instantiater with a success threshold of
$10^{-10}$. The default BQSKit partitioner handled all circuit partitioning.

Similarly to the original SABRE evaluation, we configured PAM with a
maximum extended set size $|E|$ of 20 and a weight $W_E$ of 0.5. We used a decay delta
of 0.001 and reset the decay every five steps or after mapping a gate.
When discovering the initial layout, we performed two complete
forward-and-backward passes. PAM's gate count
heuristic weight $W_P$ is set to 0.1. We used the
BQSKit implementation and the same values for common parameters when
evaluating the original SABRE algorithm. For the Qiskit, BQSKit, and TKET compilers we used the recommended settings with maximum optimization level.


The experimental results are verified with classical simulation and numerical instantiation based error upper-bound verification ~\cite{quest, younis2022instantiation}. The error upper bounds on all outputs were less than $10^{-8}$. 

\section{Evaluation}
\label{sec:eval}

\begin{figure}
    \centering
    \includegraphics[width=0.9\linewidth, height=4.5 cm]{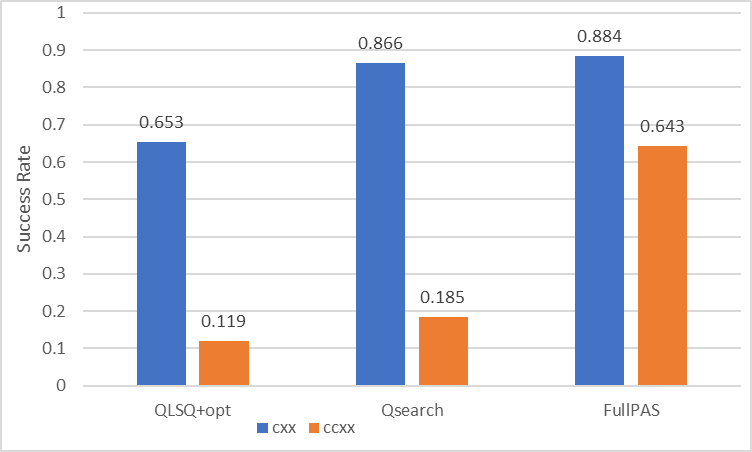}
    \caption{\footnotesize \it Comparison of OLSQ+Opt, Qsearch, and FullPAS on \texttt{ibm\_oslo}}
    \label{fig:real_oslo}
\end{figure}

\begin{table*}
 \caption{Mapping and optimizing a quantum circuit benchmark suite to a fully connected topology.}
  \label{tab:atapam}
  \centering
  \small
  \resizebox{0.8\linewidth}{0.75in}{%
  \begin{tabular}{|c|c|c|c|c|c|c|c|c|c|c|}
  \hline
    &  \multicolumn{2}{c|}{\textbf{SABRE}} &  \multicolumn{2}{c|}{\textbf{Qiskit}} &  \multicolumn{2}{c|}{\textbf{TKET}} & \multicolumn{2}{c|}{\textbf{BQSKit}} & \multicolumn{2}{c|}{\textbf{PAM3}}\\
    \hline
     \textbf{benchmark} & $\#CX$ & time(s)& $\#CX$ & time(s) & $\#CX$ & time(s) & $\#CX$ & time(s) & $\#CX$ & time(s)\\
    \hline
adder63 & 1405 & 3.23 & 1405 & 9.98 & 484 & 14.45 & 1195 & 34.41 & \textbf{442} & 187.08 \\\hline
mul60 & 11405 & 24.09 & 11403 & 72.27 & 4144 & 428.55 & 9926 & 225.75 &  \textbf{3938} &  1493.63 \\\hline
qft5 & 20 & 0.28 & 20 & 2.42 & 20 & 0.49 & 20 & 4.04 & \textbf{18} & 18.59 \\\hline
qft64 & 1880 & 3.78& 1720 & 10.74 & 1784 & 24.61 & 1771 & 188.87 &  \textbf{1665} & 771.31\\\hline
grover5 & 48 & 0.35 & 48 & 2.69 & 46 & 0.79 & 48 & 10.82 & \textbf{44} & 51.80  \\\hline
hub18 & 3541 & 6.87 & 3529 & 22.86 & \textbf{3428} & 76.35 & 3498 & 50.59 & 3459 & 524.00 \\\hline
shor26 & 21072 & 42.01 & 21072 & 109.30 & 20884 & 836.27 & 16319 & 1020.94 & \textbf{14950} & 9976.45 \\\hline
qaoa12 & 198 & 0.58 & 198 & 3.15 & 132 & 2.03 & 191 & 8.43 & \textbf{129} & 75.93\\\hline
tfim64 & 4032 &  9.79 & 4030 & 31.17 & 4032 & 107.38 & 4013 & 169.91 & \textbf{2820} & 2232.45 \\\hline
tfxy64 & 4032 & 9.84 & 4032 & 31.00 & 4032 & 108.84 & 4014 & 170.04 & \textbf{3294} & 1791.33 \\\hline
  \end{tabular}
  }
  \end{table*}

  \begin{table*}[bhtp]
  \caption{Mapping and optimizing a quantum circuit benchmark suite targeting Rigetti's Aspen M2 chip}
  \label{tab:m2pam}
  \centering
  \small
    \resizebox{0.8\linewidth}{0.75in}{%
      \begin{threeparttable}
  \begin{tabular}{|c|c|c|c|c|c|c|c|c|c|c|}
  \hline
    &  \multicolumn{2}{c|}{\textbf{SABRE}} &  \multicolumn{2}{c|}{\textbf{Qiskit}} &  \multicolumn{2}{c|}{\textbf{TKET}} & \multicolumn{2}{c|}{\textbf{BQSKit}} & \multicolumn{2}{c|}{\textbf{PAM3}}\\
    \hline
     \textbf{benchmark} & $\#CX$ & time(s)& $\#CX$ & time(s) & $\#CX$ & time(s) & $\#CX$ & time(s) & $\#CX$ & time(s)\\
    \hline
adder63 & 3931 & 6.62 & 3250 & 23.90 & 1798 & 15.51 & 3801 & 85.62 & \textbf{1566} & 301.63\\\hline
mul60 & 30386 & 38.24 & 24832 & 196.47 & 14708 & 441.12 & 25580 & 514.29 & \textbf{11172}  & 2400.01\\\hline
qft5 & 41 & 0.39 & 34 & 2.68 & 35 & 0.26 & 29 & 4.18 & \textbf{28} & 24.23\\\hline
qft64 & 6383 & 10.38 & 5107 & 34.19 & 4970 & 25.40 & 5575 & 293.78 &  \textbf{3861} & 1194.87\\\hline
grover5 & 108 & 0.49 & 110 & 3.05 & 82 & 0.57 & 63 & 13.57 & \textbf{59} & 89.74\\\hline
hub18 & 15151 & 10.58 & 13031 & 67.51 & \textbf{11680} & 77.20 & 12236 & 187.58 & 11785 & 1089.22\\\hline
shor26 & 44907 & 33.39 & 39171 & 220.17 & 46192 &  862.52 & 32110 & 795.92 & \textbf{29055} & 15528.36\\\hline
qaoa12 & 303 & 0.55 & \textbf{198} & 4.34 & 253 & 1.96 & 219 & 13.29 & 302 (\textbf{188}) & 100.12\\\hline
tfim64 & 8403 & 25.13 & \textbf{4032} & 95.63 & \textbf{4032} & 109.26 & 6040 & 170.96 & 4532 (\textbf{2804}) & 4014.57\\\hline
tfxy64 & 8403 & 24.52 & \textbf{4032} & 138.95 & \textbf{4032} & 110.24 & 7884 & 292.50 & 5963 (\textbf{3319}) & 5577.72\\\hline
  \end{tabular}
  \begin{tablenotes}
  \item  The numbers in brackets represent the experimental results of PAM3 with extra isomorphism check.
  \end{tablenotes}
  \end{threeparttable}
    }
  \end{table*}

\subsection{Block mapping}
\label{subsec:block_mapping}
We first evaluated the mapping and optimization potential for synthesis
and our permutation-aware synthesis framework at the block level.
We selected two architectures to evaluate the different methods: a line
with only nearest-neighbor connectivity and a fully connected topology.
Figures~\ref{tab:linepas} and \ref{tab:atapas} respectively detail
the final CNOT counts and total compile time for the two different
target architectures.

Fully permutation-aware synthesis (FullPAS) produced shortest circuits in
all cases. 
FullPAS built circuits with an average of 42\%,
43\%, 42\%, and 21\% fewer gates than did Qiskit, TKET, OLSQ, and QSearch, respectively, where
SeqPAS produced circuits with an average of 37\%, 37\%, 36\%, and 12\% fewer
gates.

An optimal decomposition is not always precomputed and
available or trivial to compute by hand, however, as in the case of the
controlled MS gates. FullPAS resulted in a cxx circuit with 19\% and 24\% of the  gates in the best nonsynthesized
result when compiling to the linear or fully connected topology,
respectively. This improvement is even more pronounced in the case of the
ccxx circuit, where FullPAS produced circuits with as much as 27 times fewer
gates; however, improvements over Qsearch are much more modest.
Nonetheless, these modest gains are still significant. FullPAS compiled
a 5-CNOT qft3 circuit for all topologies; this is, to the best of our
knowledge, the new best-known implementation of this essential circuit.

These significant improvements in quality require many synthesis calls
and, as a result, more runtime than other methods require. Since FullPAS
calls for synthesizing all pairs of input and output permutations, its
scaling is limited. 
SeqPAS, however, is much more palatable, with an average runtime of 24.25 seconds
for 3-qubit blocks and 3175 seconds for 4-qubit blocks.

We evaluate the cxx and ccxx benchmarks on a 27-qubit \texttt{ibm\_oslo} computer. The gate counts are reported in Figure~\ref{tab:linepas}. As shown in Figure~\ref{fig:real_oslo}, FullPAS generates the circuit that has the highest success rate. In the ccxx example, permutation-aware synthesis reduces the CNOT gate count from 15 to 10, resulting in a 3.5x success rate boost.


  \begin{table*}[bhtp]
  \caption{Mapping and optimizing a quantum circuit benchmark suite targeting Google's Bristlecone chip}
  \label{tab:bristleconepam}
  \centering
  \small
    \resizebox{0.8\linewidth}{0.75in}{%
  \begin{tabular}{|c|c|c|c|c|c|c|c|c|c|c|c|}
  \hline
    &  \multicolumn{2}{c|}{\textbf{SABRE}} &  \multicolumn{2}{c|}{\textbf{Qiskit}} &  \multicolumn{2}{c|}{\textbf{TKET}} & \multicolumn{2}{c|}{\textbf{BQSKit}} & \multicolumn{2}{c|}{\textbf{PAM3}}\\
    \hline
     \textbf{benchmark} & $\#CX$ & time(s)& $\#CX$ & time(s) & $\#CX$ & time(s) & $\#CX$ & time(s) & $\#CX$ & time(s)\\
    \hline
adder63 & 3274 & 6.79 & 2726 & 21.22 & 1326 & 15.89 & 2755 & 65.79 & \textbf{925} & 297.99\\\hline
mul60 & 24974 & 32.35 & 20014 & 171.35 & 11989 & 437.34 & 18396 & 361.08 &  \textbf{9169} & 2404.43\\\hline
qft5 & 35 & 0.30 & 30 & 2.54 & 32 & 0.26 & 31 & 3.49 & \textbf{22} & 25.65\\\hline
qft64 & 5153 & 9.14 & 4304 & 30.27 & 4175 & 25.30 & 4262 & 228.93 &  \textbf{3624} & 1195.86\\\hline
grover5 & 108 & 0.36 & 96 & 3.03 & 82 & 0.58 & 85 & 3.85 &  \textbf{62} & 81.08\\\hline
hub18 & 11227 & 8.89 & 10137 & 55.49 & 9084 & 77.05 & 9064 & 124.58 & \textbf{8682} & 1095.16\\\hline
shor26 & 38241 & 29.86 & 36365 & 204.72 & 38070 & 849.95 & 28624 & 659.49 & \textbf{24021} & 15547.82\\\hline
qaoa12 & 198 & 0.36 & 198 & 64.47 & 237 & 1.94 & \textbf{205} & 12.03 & 243 (\textbf{188}) & 95.88\\\hline
tfim64 & 6591 & 24.35& 4828 & 80.56 & 4773 & 156.35 & 5187 & 173.95 & \textbf{4344} (\textbf{2804}) & 4312.13\\\hline
tfxy64 & 6591 & 24.13 & 5204 & 78.47 & \textbf{4773} & 158.04 & 5814 & 255.63 & 4778 (\textbf{3319}) & 5825.42\\\hline
  \end{tabular}
  }
  \end{table*}

    \begin{table*}[bhtp]
  \caption{Mapping and optimizing a quantum circuit benchmark suite targeting IBM's Eagle chip}
  \label{tab:eaglepam}
  \centering
  \small
    \resizebox{0.8\linewidth}{0.75in}{%
  \begin{tabular}{|c|c|c|c|c|c|c|c|c|c|c|}
  \hline
    &  \multicolumn{2}{c|}{\textbf{SABRE}} &  \multicolumn{2}{c|}{\textbf{Qiskit}} &  \multicolumn{2}{c|}{\textbf{TKET}} & \multicolumn{2}{c|}{\textbf{BQSKit}} & \multicolumn{2}{c|}{\textbf{PAM3}}\\
    \hline
     \textbf{benchmark} & $\#CX$ & time(s)& $\#CX$ & time(s) & $\#CX$ & time(s) & $\#CX$ & time(s) & $\#CX$ & time(s)\\
    \hline
adder63 & 4906 & 9.08 & 4172 & 34.09 & 2318 & 16.02 & 4070 & 107.74 & \textbf{1827} & 316.22\\\hline
mul60 & 37982 & 44.92 & 31284 & 349.58 & 18000 & 442.01 & 30817 & 612.56 & \textbf{14553}  & 2493.17\\\hline
qft5 & 41 & 1.37 & 35 & 2.72 & 38 & 0.26 & 32 & 6.26 & \textbf{28} & 60.02\\\hline
qft64 & 6491 & 11.31 & 5760 & 46.52 & 5682 & 26.00 & 5511 & 321.04 &  \textbf{4466} & 1190.36\\\hline
grover5 & 114 & 1.38 & 122 & 3.14 & 82 & 0.58 & 60 & 12.00 & \textbf{59} & 79.73\\\hline
hub18 & 17692 & 12.84 & 16990 & 93.66 & \textbf{13648} & 77.91 & 14288 & 222.41 & 14365 & 1161.29\\\hline
shor26 & 50334 & 40.52 & 43705 & 239.15 & 54156 & 858.67 & 35659 & 978.47 & \textbf{34205} &  15684.63\\\hline
qaoa12 & 309 & 1.54 & \textbf{198} & 3.38 & 241 & 1.81 & 276 & 12.60 & 232 (\textbf{188}) & 96.14\\\hline
tfim64 & 12126 & 41.26 & \textbf{4032} & 402.50 & \textbf{4032} & 107.16 & 8730 & 241.97 & 10652 (\textbf{2804}) & 4493.52\\\hline
tfxy64 & 12126 & 40.90 & \textbf{4032} & 395.86 & \textbf{4032} & 108.33 & 9469 & 297.04 & 8260 (\textbf{3319}) & 5852.39\\\hline
  \end{tabular}
  }
  \end{table*}

    \begin{table*}[bhtp]
  \caption{Quality of solutions and compile time (s) of OLSQ + opt and PAM3}
  \label{table:olsq_comparison}
  \centering
  \small
    \resizebox{0.8\linewidth}{!}{%
  \begin{tabular}{|c|c|c|c|c|c|c|c|c|c|c|c|c|}
  \hline    
&  \multicolumn{4}{c|}{\textbf{Fully-connected}} &  \multicolumn{4}{c|}{\textbf{Aspen M2}} &  \multicolumn{4}{c|}{\textbf{IBM Eagle}}\\
  \hline
    &  \multicolumn{2}{c|}{\textbf{OLSQ}} &  \multicolumn{2}{c|}{\textbf{PAM3}}&  \multicolumn{2}{c|}{\textbf{OLSQ}} &  \multicolumn{2}{c|}{\textbf{PAM3}}&  \multicolumn{2}{c|}{\textbf{OLSQ}} &  \multicolumn{2}{c|}{\textbf{PAM3}} \\
    \hline
     \textbf{benchmark} & $\#CX$ & time(s)& $\#CX$ & time(s) & $\#CX$ & time(s)& $\#CX$ & time(s) & $\#CX$ & time(s)& $\#CX$ & time(s)  \\
    \hline
    alu-v0 & 17 & 2.8& \textbf{13} & 15.17 &28 & 207.36 & \textbf{21} & 27.89 & 28 & 324.10 & \textbf{21} & 27.99\\\hline
    qft5 & 20 & 1.58 & \textbf{18} & 18.59 & \textbf{28} & 12.20 & \textbf{28} & 24.23 & \textbf{28} & 11.86 & \textbf{28} & 60.02\\\hline
    grover5 & 48 & 2.25 & \textbf{44} & 51.80 & 76 & 393.67 & \textbf{59} & 89.74 & 76 & 352.52 & \textbf{59} & 79.73\\\hline
    qaoa8 & 24 & 1.98 & \textbf{23} & 8.1 & 38 & 66.18 & \textbf{35} & 9.33 & \textbf{45} & 666.38 & 47 & 11.75\\\hline
  \end{tabular}
  }
  \end{table*}

\subsection{Large circuits}

To evaluate the mapping methods, we chose four real quantum architectures
implemented  in state-of-the-art quantum processors: Rigetti's Aspen
M2 80-qubit chip~\cite{Rigetti_m2}, Google's 72-qubit Bristlecone
chip~\cite{Google_bristlecone}, IBM 127-qubit Eagle chip~\cite{ibm_eagle}, and
a 64-qubit fully-connected topology similar to trapped-ion
architectures~\cite{ionq, quantinuum}.

The 3-qubit version of the PAM algorithm (PAM3) produced the shortest circuits
in the most trials, with an average of 35\%, 18\%, 9\%, and 21\%
fewer gates than SABRE, Qiskit, TKET, and BQSKit. The results are demonstrated in Table~\ref{tab:atapam},\ref{tab:m2pam},\ref{tab:bristleconepam},\ref{tab:eaglepam}. OLSQ cannot find any solution for the benchmarks with tens of qubits. Therefore we exclude it in the large circuit comparison. PAM3 built
the shortest circuit in 29 out of the 40 trials (10 circuits and 4
architectures). Three out of the eleven cases where PAM3 was not the shortest were 
 with the 18-qubit Hubbard, where PAM3 built circuits with 0.9\%, 0.9\% and
5.2\% more gates than TKET did. By adding isomorphism check, PAM3 produces the shortest circuit in 37 trials. The data with isomorphism check is presented in brackets. 

\textit{QAOA, TFIM, and TFXY:} In the remaining eight times PAM3 produced a
longer circuit,  the cases invovled QAOA, TFIM, or TFXY circuits. This result is due to placement. These three circuits all require only linear connections.
Theoretically, they can be mapped to all four chips without routing.
Qiskit and TKET do a subgraph isomorphism check, which sometimes catches
a perfect placement. This extra check highlights the downside of comparing
our experimental mapping algorithm with complete commercial compilers with
max settings. A lot of extra bells and whistles  can divert
compilation flow in specific cases. However, in the cases where they
did not catch the isomorphism check, we produced shorter circuits. Additionally,
if we implement the same isomorphism check, we can outperform them because
we can often further reduce the circuit depth on a line. For example,
suppose we pick a perfect placement and map the QAOA to a line with PAM3.
In that case, we get a result with 188 CNOTs, which can be directly placed
on any of the four experiment architectures and is shorter than all other
compilers' output. Similarly, TFIM and TFXY can be compiled with 2804 and 3319 CNOTs by adding the isomorphism check. The isomorphism check only takes tens of seconds which is negligible compared to PAM's compilation time. Moreover,  PAM3 gets good placement when
compiling the TFIM circuit to the Bristlecone architecture and produces a result
with fewest CNOTs.

\subsection{Comparison with optimal layout solver}
In this section, we compare the solution quality and compilation time of PAM with the optimal solver OLSQ to evidence the effectiveness of permutation-aware mapping. Table~\ref{table:olsq_comparison} demonstrates the final gate count and the compile time. We use OLSQ for routing followed by Qiskit optimizations. OLSQ finds the optimal mapping and routing that minimizes the number of inserted SWAP gates. However, since PAM directly synthesizes the unitary based on hardware connectivity, the resulting circuit is on average 10.7\% smaller than OLSQ. Moreover, the optimal solver have scalability issue, it cannot find any solution on the coupling map of Google's Bristlecone. As shown in the table, when compiled with limited backend connectivity(Aspen-M2, IBM-Eagle), PAM has shorter compilation time than OLSQ for most benchmarks.
  
\subsection{Scaling beyond the NISQ era}

\begin{figure}
    \centering
   \includegraphics[width=0.9\linewidth, height=45mm] {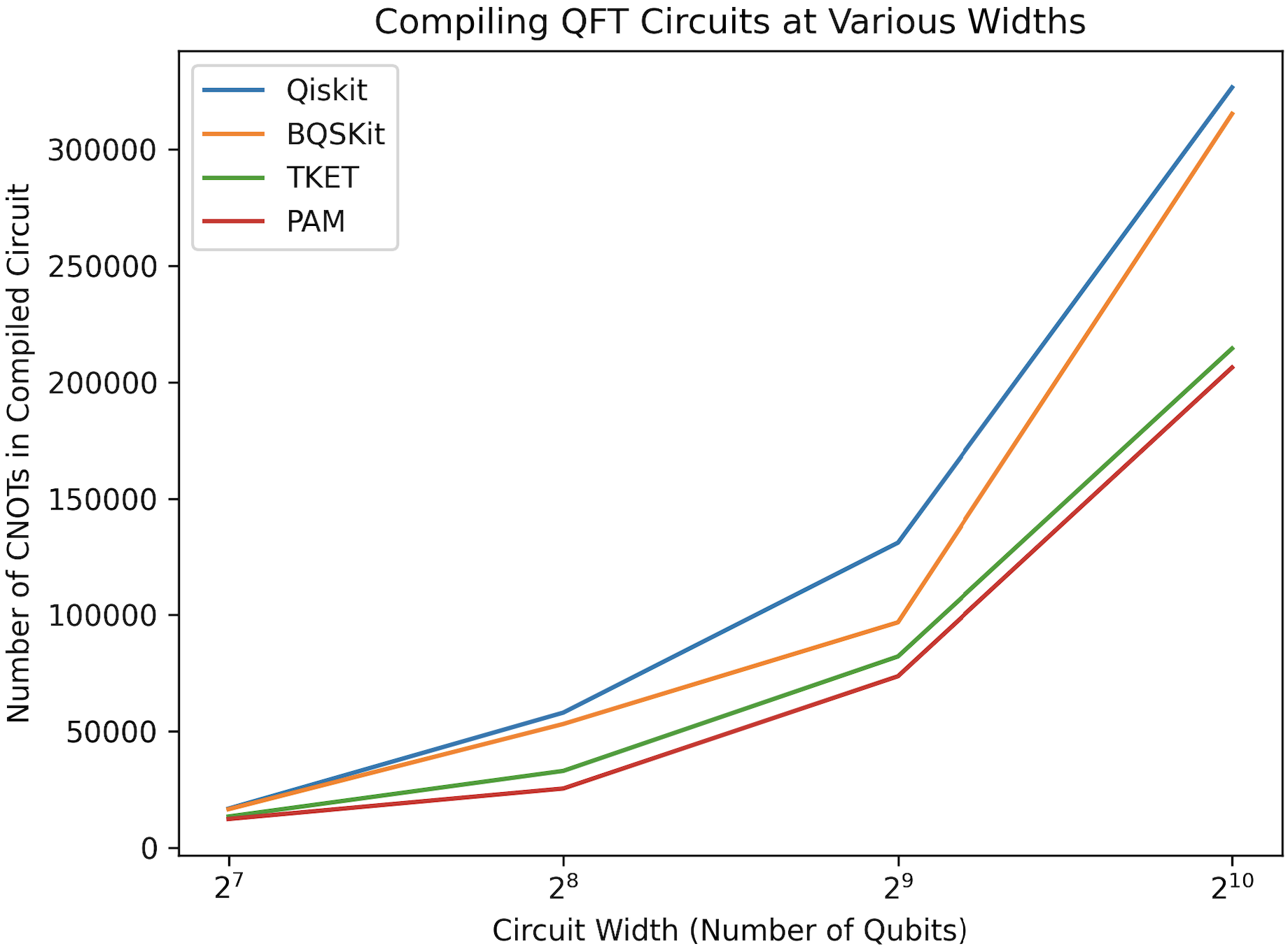}
    \includegraphics[width=0.9\linewidth, height=45mm]{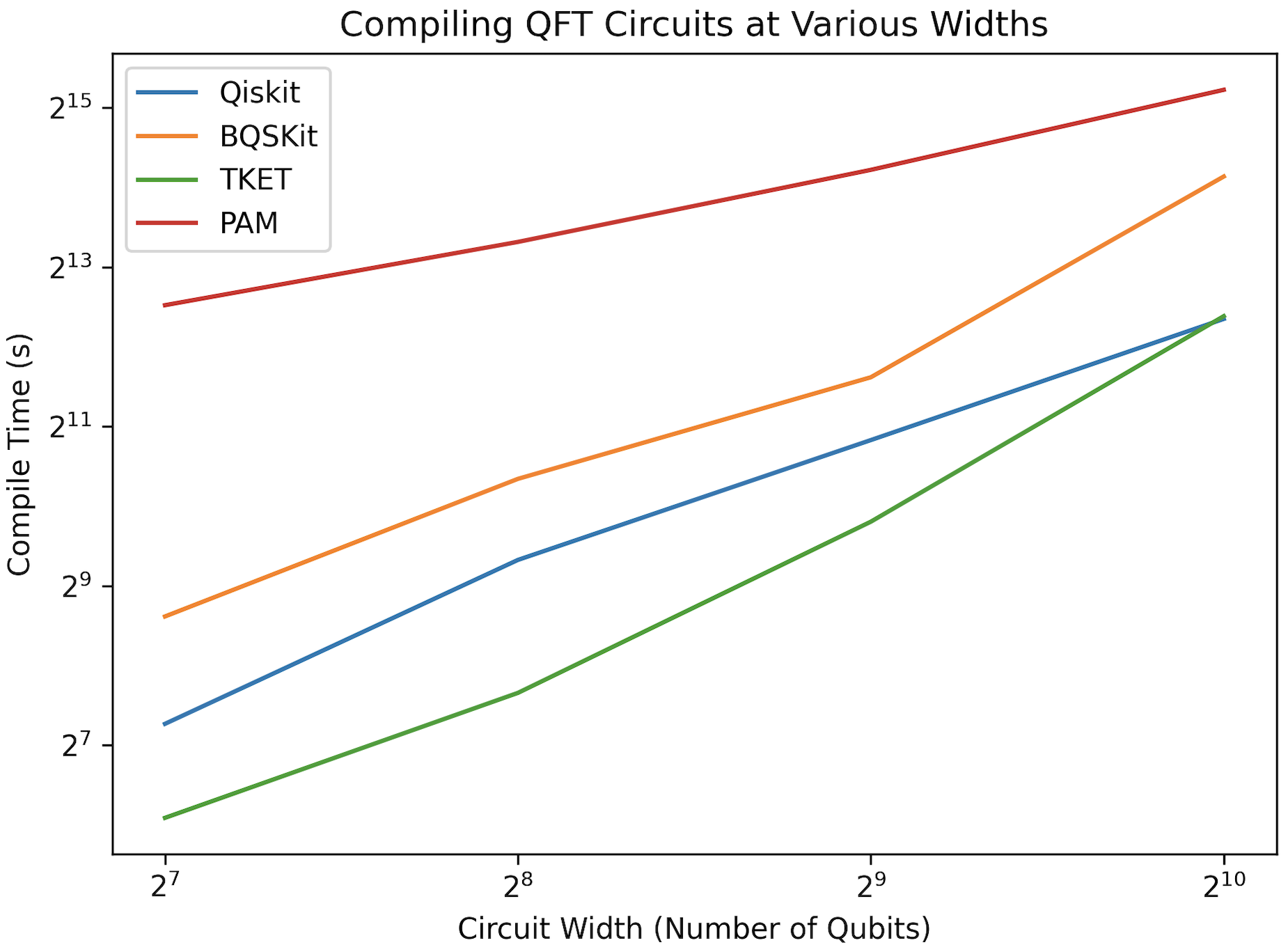}
    \caption{\footnotesize \it
        Scaling of the QFT benchmark}
    \label{fig:qft_scaling}
\end{figure}

To evaluate the scalability of the mapping algorithms past the
capabilities of quantum hardware today, we generated a set of QFT circuits ranging from 128 qubits to 1024 qubits and mapped them to a proposed heavy-hexagonal chiplet architecture~\cite{heavy_hex_chiplet}. We built an architecture following the tree-of-grids approach with a 3-node tree containing a $4\times4$-grid of 27-qubit chiplets. The results are shown in Figure~\ref{fig:qft_scaling}. As the number of qubits increases, PAM always generates the circuit with the fewest gate count, and the gaps between the compilation time of PAM and other compilers are narrowing. This highlights the scalability of our routing framework and the capability to handle future hardware designs. For the 1024 qubit QFT algorithm, PAM generates the smallest circuit with 206310 CNOTs, with 8159 CNOT gate reduction compared to the next best result from TKET.


\subsection{Closer examination of the improvements}

\begin{figure}
    \centering
    \includegraphics[width=0.9\linewidth, height =4.5 cm]{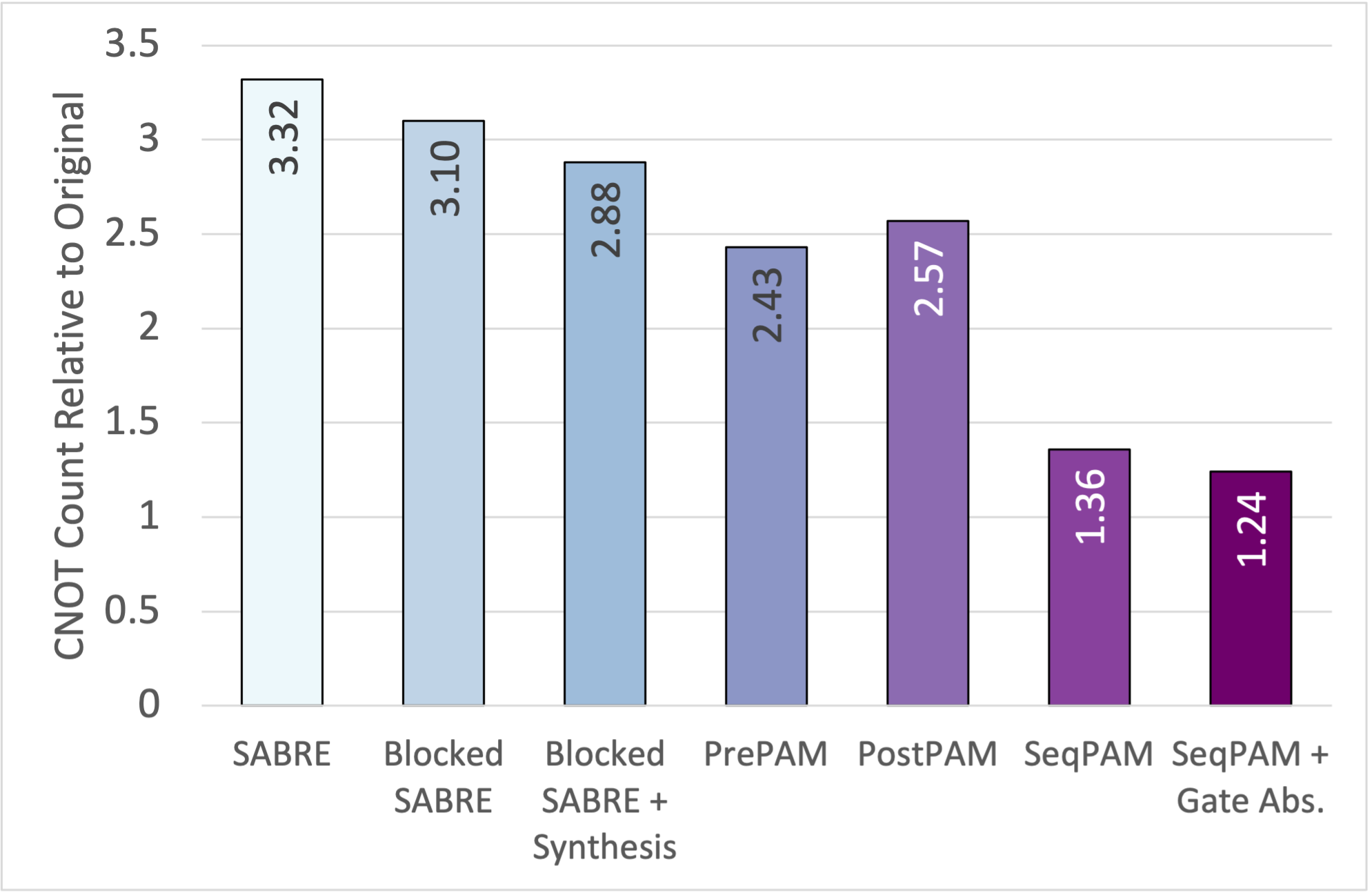}
    \caption{\footnotesize \it
        A breakdown of the improvements each individual feature adds
        on top of the SABRE algorithm. These results
        are from compiling the 60-qubit multiply circuit to the
        IBM Eagle architecture.
    }
    \label{fig:breakdown}
\end{figure}

Since we have introduced a few features that improve upon the original
SABRE algorithm, we thought it necessary to analyze how much
each improves individually. In Figure~\ref{fig:breakdown} each additional
feature is measured separately when compiling
the multiply circuit. The PrePAM and PostPAM represent the cases where we only enable permutation on the input or output sides. We start with the original SABRE algorithm and then
introduce the concept of partitioning. Just by mapping blocks in a circuit rather than gates, we can see an improvement which we believe is because this increases the lookahead factor of the SABRE. 
Using synthesis to route
inside the blocks improves the results. When we introduce the concept of permutation-aware-synthesis,
we see the next big jump even if it is just one-sided with PrePAM and
PostPAM. Furthermore, doing both sides in SeqPAM introduces the biggest
jump. Repartition and resynthesis (Gate absorption) further improve the result in the 3-qubit case.

\section{Discussion}
\label{sec:disc}

\subsection{Relevance to trapped ions}

We have mentioned that our permutation-aware algorithms can leverage hardware connectivity by design. This effect is
visible when compiling the linearly connected tfim64 and tfxy64 circuits
to the fully connected topology. No other compiler can effectively utilize the full-connectivity by design; however,
PAM3 produces a circuit with 2,820 CNOTs versus the 4,032 tfim64 input. The
next best is BQSKit with 4,013 CNOTs. These TFIM input circuits were
previously the best-known implementations of these real-time evolution
circuits.

One way to quantify this concept is by using Supermarq's~\cite{supermarq}
program communication metric. The metric measures
how sparsely or densely a circuit's logical connectivity is. A program
communication value of 0 implies no connectivity, while a value of 1
implies that every qubit requires a connection with every other qubit.
The 12-qubit QAOA started with a communication score of 0.167 but ended
with a score of 1. This shift implies that we took the linearly connected
input and returned a fully connected output with fewer CNOTs than any
other compiler. Additionally, the scores improved in all the other cases
when compiling to an all-to-all architecture and in most cases with the
densely connected Bristlecone architecture. Increasing program communincation has particular significance for trapped-ion
architectures. This class of quantum processors allows a program to apply
a gate to any two pairs of qubits. PAM's ability to fully leverage the hardware connectivity is advantageous as an optimization pass for these architectures.


\subsection{Building PAM into a workflow}

PAM3 produced circuits shorter than state-of-the-art compilers in many
trials tested; however, PAM3 is just a mapping algorithm with good
optimization potential. We can replace the mapping algorithm inside
Qiskit, TKET, and BQSKit and sum up to a better compiler. We did this
and compiled the qft64 to the M2 chip and saw an additional reduction
of 15\%, 55\%, and 13\% CNOTs when compiling with Qiskit, TKET, and BQSKit, respectively.

\subsection{Tunability}

PAM3 has built  efficient circuits, but it always tends to take a lot
more time than other compilers. Algorithm scientists will spend the time
necessary to produce the best circuit possible, mainly since circuits are
often compiled only once, quantum computer time is  expensive, and
longer circuits are more likely to produce erroneous results. Our proposed algorithm has many parameters one can adjust to
improve runtime. In particular, the number of multistarts for instantiation
has the most significant impact on runtime. 
For example, if we decrease the number of multistart to one, the runtime of the the shor26 reduces from 
15547.82 seconds to 5908.40 seconds for the Bristlecone architecture. 



\section{Conclusion}
\label{sec:conc}

In this work we built on top of both general unitary synthesis and
heuristic-based mapping algorithms by introducing the idea of
permutation awareness with respect to the mapping problem. This codesign
was accomplished by first lifting mapping from the native gate level to
the block level. This elevation led to generally good results on its
own but also introduced many new opportunities for optimization. While we have shown that these algorithms are effective and
competitive, we have demonstrated the ability to leverage hardware connectivity is particularly helpful for optimizing the circuits for fully connected architectures. We have also shown
the implementability and tunability of our algorithms with potential
application in existing compiler frameworks.

\section*{Acknowledgements}
This work was supported by the DOE under contract DE-5AC02-05CH11231 and DE-AC02-06CH11357, through the Office of Advanced Scientific Computing Research (ASCR) Quantum Algorithms Team and Accelerated Research in Quantum Computing programs.


\bibliographystyle{IEEEtranS}
\bibliography{quantum}

\end{document}